\documentclass[apj,a4paper,12pt,useAMS]{emulateapj}
\usepackage{amsmath}
\usepackage{cases}
\usepackage{amssymb}
\usepackage{graphicx}

\begin{document}
\title{The evolution of a newborn millisecond magnetar with a propeller-recycling disk}
\author{Shao-Ze Li$^{1,2}$, Yun-Wei Yu$^{2}$, He Gao$^{1}$ \& Bing Zhang$^{3}$}

\altaffiltext{1}{Department of Astronomy, Beijing Normal University,
Beijing 100875, China; gaohe@bnu.edu.cn}
\altaffiltext{2}{Institute of Astrophysics,
Central China Normal University, Wuhan 430079, China; yuyw@mail.ccnu.edu.cn}
\altaffiltext{3}{Department of Physics and Astronomy, University of
Nevada Las Vegas, NV 89154, USA}

\begin{abstract}
A rapidly rotating and highly magnetized neutron star (NS) could be
formed from the explosive phenomena such as superluminous supernovae
and gamma-ray bursts. This newborn NS can substantially influence
 the emission of these explosive transients through its spin-down. The
spin-down evolution of the NS can sometimes be affected by fallback
accretion, although it is usually regulated by the magnetic dipole
radiation and gravitational wave radiation of the NS. Under
appropriate conditions, the accreting material can be firstly
ejected and subsequently recycled back, so that the accretion disk
can keep in a quasi-steady state for a long time. Here we describe
the interaction of the NS with such a propeller-recycling disk and
their co-evolution. Our result shows that, the spin-down of the NS
can be initially dominated by the propeller, which prevents the disk
material from falling onto the NS until hundreds or thousands of
seconds later. It is suggested that the abrupt fall of the disk
material onto the NS could significantly suppress the magnetic
dipole radiation and then convert the NS from a normal magnetar to a
low-field magnetar. This evolution behavior of the newborn NS can
help to understand the very different influence of the NS on the
early GRB afterglows and the late supernova/kilonova emission.
\end{abstract}

\section{introduction}

It is widely considered that a rapidly rotating and highly
magnetized neutron star (NS) could be formed from the core-collapse
of some massive stars, which can be characterized by a long-duration
gamma-ray burst (GRB) or a superluminous supernova (SLSN). The
existence of such a millisecond magnetar is beneficial for
explaining the long-lasting energy engine of these phenomena for the
first few hours or days (Dai \& Lu 1998a,b; Zhang \&
M\'{e}sz\'{a}ros 2001; Kasen \& Bildsten 2010; Yu et al. 2017). It
was also suggested that a millisecond magnetar can be formed from
some mergers of double NSs which produce the short-duration GRBs
(Dai et al. 2006; Rowlinson et al. 2013; Zhang 2013; Yu et al. 2013;
Metzger \& Piro 2014). This hypothesis, if true, can provide a
robust constraint on the equation of state of NS matter, since the
remnant NS of mergers must have very high masses. Therefore, in
order to confront this model with observations in more details in
the future, it is necessary to investigate the energy release
process of these millisecond magnetars more carefully. In addition,
such an investigation can also help to discover the theoretically
predicted accretion-induced collapse events (AICs) of white dwarfs
in future transient surveys (Dessart et al. 2006; Piro \& Thompson
2014; Yu et al. 2019a,b).

The influence of a newborn NS on the accompanied transient emission is
highly dependent on the spin-down behavior of the NS. The spin-down is
usually determined by the magnetic dipole (MD) radiation and,
sometimes, also by the gravitational wave (GW) radiation if the
ellipticity of the NS is high enough. Meanwhile, it is natural to
consider that a fraction of the material in supernova ejecta or
merger ejecta could not escape from the central NS and finally fall back onto the NS. In this case, the
early evolution of the newborn NS and consequently the transient
emission would be influenced by this fallback accretion (Dai \& Liu
2012; Kumar et al. 2008a,b), despite the majority of the material
falls only on a very short timescale. However, it should still be
noticed that the accretion can sometimes enter a propeller state, if
the NS is highly magnetized and rapidly rotating, as expected for
the NSs in GRBs and SLSNe. In this case, the material could be ejected
away by the centrifugal force rather than be accreted onto the NS
(Illarionov \& Syunyaev 1975; Alpar \& Shaham 1985; Lovelace et al.
1999). This propeller effect has been widely discussed in the
literature and supported by the observations of some accreting
millisecond pulsars (van der Klis et al. 2000; Patruno et al. 2009;
Patruno \& D'Angelo 2013; Bult \& van der Klis 2014). Piro \& Ott
(2011) suggested that some unusual supernovae could be powered by a
propeller outflow, in addition to the traditional radioactive power.
Gompertz et al. (2014) used the propeller outflow to explain the
extended emission and the X-ray afterglow plateau of some short
GRBs. The combining effects of fallback accretion and propeller on
GRBs and SLSNe have been investigated by Metzger et al. (2018).

However, alternatively, it is still worth considering that the
propeller, even though operating, may not be strong enough to eject
the material to infinity. Instead, the propeller-ejected material
could finally fall back to the disk at a certain radius. In other
words, the material in the disk can be recycled for a long time and
maintain a quasi-steady state, which is somewhat similar to a dead disk surrounding a pulsar in a binary (Syunyaev \& Shakura 1977; Spruit \&
Taam 1993; D'Angelo \& Spruit 2010, 2012; Romanova et al. 2018). Nevertheless, even though the disk is quasi-steady, the NS can still be
spun down by the propeller and, finally, the disk material must fall
onto the NS. When the material lands on the polar surface of the NS,
it can in principle repel the magnetic field lines from the polar
cap region or even bury some field lines. As a result of this
delayed accretion, the effective dipolar magnetic field can be
decreased and then the MD radiation is suppressed, which can undoubtedly influence the energy
release from the NS and the accompanied transient emission. The
purpose of this paper is to calculate the early evolution of a
newborn millisecond magnetar under the influence of such a propeller-recycling accretion disk.

\section{The accretion and propeller}
The primary reason for the existence of a long-lived
propeller-recycling disk is that the high magnetic filed of the
central NS can drag and accelerate the disk material and then
prevent the material from falling onto the NS. Therefore, balancing
the magnetic pressure and the ram pressure of the accretion flow,
the inner boundary of the disk can be roughly determined by the
Alfv\'{e}n radius as\footnote{Elaborated calculations show that the
proportional coefficient between the inner disk radius and the
Alfv\'{e}n radius could range from 0.5 to 1 (Ghosh \& Lamb 1979a,b;
Arons 1986, 1993), which however essentially cannot affect the
calculations in this paper.}
\begin{eqnarray}
R_{\rm m}&=& {\left(\mu^{4}\over GM_{\rm s}
\dot{M}_{\rm disk}^{2}\right)}^{1/7}\nonumber\\
&=&2.8\times10^{6}M_{\rm s,1.4}^{-1/7}R_{\rm s,6}^{12/7}B_{\rm
s,14}^{4/7}\dot{M}_{\rm disk,-5}^{-1/7}\rm \,cm,\label{Rm}
\end{eqnarray}
where $\mu=B_{\rm s}$$R_{\rm s}^{3}$ is the magnetic moment of NS,
$G$ is the gravitational constant, $\dot{M}_{\rm disk}$ is the mass
rate of accretion flow in the disk, $B_{\rm s}$, $R_{\rm s}$ and
$M_{\rm s}$ are the strength of surface dipolar magnetic field, the
radius, and the mass of NS, respectively. Hereafter the conventional
notation $Q_{x}=Q/10^x$ is adopted in the cgs units and,
additionally, $M_{\rm s,1.4}=M_{\rm s}/1.4\rm M_{\odot}$ and
$\dot{M}_{\rm disk,-5}=\dot{M}_{\rm disk}/10^{-5}\rm
M_{\odot}~s^{-1}$. Once the disk material enters the region inside
$R_{\rm m}$, the high magnetic pressure can enforce the material to
corotate with the NS. Then, we can judge whether the disk is in the
accretion or the propeller state by comparing this Alfv\'{e}n radius
with the corotation radius. Here the corotation radius can be
defined by
\begin{eqnarray}
R_{\rm c}&=&\left({GM_{\rm s}\over\Omega_{\rm
s}^{2}}\right)^{1/3}=1.7\times10^{6}M_{\rm s,1.4}^{1/3}P_{\rm
s,-3}^{2/3}\rm\,cm,\label{Rc}
\end{eqnarray}
where $\Omega_{\rm s}$ is the spin frequency of NS and $P_{\rm
s}=2\pi/\Omega_{\rm s}$ is the spin period. The disk material is
rotating at the the Keplerian frequency $\Omega_{\rm
K}=\sqrt{GM_{\rm s}/R^3}$ at each radius. Inside $R_{\rm c}$, the
spin frequency of NS is slower than the Keplerian frequency.

If $R_{\rm m}<R_{\rm c}$, the magnetic field would first slow down
the disk material inside $R_{\rm c}$. Once the angular momentum of the material is lost, the disk
material can no longer keep Keplerian rotation at the orbit and
finally fall onto the surface of the NS along the magnetic field
lines. Consequently, the angular momentum carried by the disk
material can be gradually transferred to the NS and spin it up. The
torque acting on the NS due to accretion can be expressed by
\begin{eqnarray}
T_{\rm acc}&=&\dot{M}_{\rm acc}R_{\rm m}^2\Omega_{\rm K,m}=\dot{M}_{\rm acc}\sqrt{GM_{\rm s}R_{\rm m}}, \label{Taccretion}
\end{eqnarray}
where $\Omega_{\rm K,m}=\sqrt{GM_{\rm s}/R_{\rm m}^3}$ is the
Keplerian frequency at the Alfv\'{e}n radius and $\dot{M}_{\rm acc}$
is the mass rate accreted onto the NS. On the contrary, if $R_{\rm
m}>R_{\rm c}$, the disk material would be accelerated by the
magnetic field to approach corotation outside $R_{\rm c}$. As a
result, the velocity of disk material exceeds the Keplerian velocity
and the gravity becomes too small to support it. Then, the
centrifugal force can strongly throw the material away from the disk
to produce a propeller outflow. Meanwhile, the NS is spun down and
the corresponding torque can be written as (Lai 2014)
\begin{eqnarray}
T_{\rm pro}&=&-\dot{M}_{\rm pro}R_{\rm m}^2\Omega_{\rm s},
\label{Tpropeller}
\end{eqnarray}
where $\dot{M}_{\rm
pro}$ is the mass rate of the propeller outflow.

In order to describe the continuous transition between the propeller
and accretion phases, a fastness parameter is defined for the NS and
disk system as
\begin{eqnarray}
\omega&=&{\Omega_{\rm s}\over \Omega_{\rm K,m}}=\left({R_{\rm m}\over R_{\rm c}}\right)^{3/2}\nonumber\\
&=&2.2M_{\rm s,1.4}^{-5/7}R_{\rm s,6}^{18/7}B_{\rm
s,14}^{6/7}\dot{M}_{\rm disk,-5}^{-3/7}P_{\rm s,-3}^{-1}.\label{omega}
\end{eqnarray}
Obviously, the value of $\omega$ reflects the proportional
relationship between the two characteristic radii. By using this
parameter, we can phenomenologically express the disk mass change
rate in terms of accretion rate and propeller mass loss rate by
\begin{eqnarray}
\dot{M}_{\rm disk}=\dot{M}_{\rm acc}+\dot{M}_{\rm pro}
\end{eqnarray}
with
\begin{eqnarray}
\dot{M}_{\rm acc}&\equiv&{1\over 1+\omega^{n}}\dot{M}_{\rm disk} \label{Maccretion}
\end{eqnarray}
and
\begin{eqnarray}
 \dot{M}_{\rm pro}&\equiv&{\omega^{n} \over
1+\omega^{n}}\dot{M}_{\rm disk},\label{Mpropeller}
\end{eqnarray}
where the artificial parameter $n>1$ is introduced to represent the
sharpness of the phase transition. The above expressions can
naturally give $\dot{M}_{\rm acc}\approx\dot{M}_{\rm disk}$ and
$\dot{M}_{\rm pro}\approx0$ for $\omega\ll 1$, and $\dot{M}_{\rm
pro}\approx\dot{M}_{\rm disk}$ and $\dot{M}_{\rm acc}\approx0$ for
$\omega\gg 1$. Substituting Equations (\ref{Maccretion}) and
(\ref{Mpropeller}) into Equations (\ref{Taccretion}) and
(\ref{Tpropeller}), the total torque acting on the NS by the disk
can be written as
\begin{eqnarray}
T_{\rm disk} = T_{\rm acc} + T_{\rm pro} =  {1- \omega^{n+1} \over
1+\omega^{n}} \dot{M}_{\rm disk} \sqrt{GM_{\rm s}R_{\rm m}}.
\end{eqnarray}
When $\omega=1$, the accretion and the propeller effects become
comparable to each other and then the total torque vanishes.

The propeller outflow escaping from the Alfv\'{e}n radius is usually
supposed to be able to escape to infinity, as long as the
centrifugal velocity of the material exceeds the escaping velocity
at the radius, i.e., $v_{\rm pro}\approx \Omega_{\rm s}R_{\rm
m}>v_{\rm esp}=\sqrt{2GM_{\rm s}/R_{\rm m}}=\sqrt{2}\Omega_{\rm
K}R_{\rm m}$. However, this assumption is actually challenged by the
following considerations. Firstly, the propeller acceleration could
be inefficient, because the material can in principle be thrown out
much before it completely achieves corotation at the Alfv\'{e}n
radius (Syunyaev \& Shakura 1977). Secondly, the kinetic energy of
the propeller material can be substantially consumed by some
internal dissipations in the outflow. Finally, the propeller
velocity is probably dominated by the azimuthal component and the
component perpendicular to the disk could be relatively small. So,
the propeller material is in principle easy to collide with the
successive falling material. The viscous friction between the
outward and inward flows can significantly decelerate the propeller
outflow. In an extreme case, the propeller outflow can even be
completely obstructed and absorbed by the accretion disk. Due to
this material recycling, the disk can basically keep in a
quasi-steady state. The net consequence of the propeller effect is
just to transfer angular momentum from the NS to the disk (Matt et
al. 2010; D'Angelo \& Spruit 2010, 2012). Therefore, the necessary
condition for the existence of a quasi-steady disk is that the disk
should initially be in a propeller state for $\omega_{\rm }>1$.
According to equation (\ref{omega}), the corresponding requirement
on the initial mass flow rate of the disk can be written as
\begin{eqnarray}
\dot{M}_{\rm disk,i}<6.0\times10^{-5}M_{\rm s,1.4}^{-5/3}R_{\rm
s,6}^6B_{\rm s,14}^2P_{\rm s,-3}^{-7/3}\rm
M_{\odot}\,s^{-1}.\label{Mdotsuit2}
\end{eqnarray}
As shown, for given NS parameters, a relatively small mass flow rate
is helpful for the appearance of a long-lived quasi-steady disk.
Therefore, qualitatively, such a disk tends to exist in the events
of double NS mergers and white dwarf AICs, in contrast to the normal
core-collapse supernovae (Macfadyen et al. 2001; Rosswog 2007).

\section{The NS and disk evolutions}
In the presence of a quasi-steady propeller-recycling disk, the spin evolution of the central NS should be determined by
\begin{eqnarray}
{dJ_{\rm s}\over dt}&=&T_{\rm md}+
T_{\rm gw}+T_{\rm disk},\label{NSangul}
\end{eqnarray}
where $J_{\rm s}=I_{\rm s}\Omega_{\rm s}$ is the angular momentum of
the NS, $I_{\rm s}\approx(2/5)M_{\rm s}R_{\rm s}^2$ is the inertial
moment of the NS, $T_{\rm md}$ and $T_{\rm gw}$ are the torques due
to MD radiation and GW radiation, respectively. Since the inner
boundary of the disk can extend into the light cylinder of the NS,
some closed magnetic field lines of the NS can be truncated in the
disk and become open lines. In this case, the MD radiation of the NS
can be enhanced by a factor of $(R_{\rm m}/R_{\rm lc})^{-2}$
(Parfrey et al. 2016). Therefore, the corresponding torque can be
written as (Metzger et al. 2018)
\begin{eqnarray}
 T_{\rm md}&=&-{\mu^2\Omega_{\rm s}^{3}\over 6c^{3}}\left({R_{\rm m}\over R_{\rm lc}}\right)^{2},\label{Tmd}
\end{eqnarray}
where $R_{\rm lc}=c/\Omega_{\rm s}$ is the light cylinder radius of
the NS. For a magnetar, it is usually considered that the internal
magnetic field can be much higher than the surface dipolar magnetic
field (e.g. Duncan \& Thompson 1992; Price \& Rosswog 2006). Then,
the NS could be deformed by the internal magnetic field to have an
ellipticity of $\varepsilon=10^{-4}(B_{\rm int}/10^{16}\rm G)^2$
(Usov 1992; Fan et al. 2013; Gao et al. 2016). In this case, a
secular GW radiation can be produced and the corresponding torque
can be written as (Zhang \& M\'{e}sz\'{a}ros 2001)
\begin{eqnarray}
 T_{\rm gw}&=&-{32GI_{\rm s}^{2}\varepsilon^{2}\Omega_{\rm s}^{5}\over
5 c^{5}},
\end{eqnarray}
which is usually much smaller than the torque given by Equation
(\ref{Tmd}) unless the surface dipolar magnetic field is much lower than
$\sim10^{12}$ G.

Beside the influence on the spin evolution of the NS, the disk can
also lead to a change of the mass and the surface magnetic field of
the NS, once the disk enters the accretion phase. On one hand, the
baryonic mass of the NS can evolve as ${dM_{\rm s,b}/ dt}=
\dot{M}_{\rm acc}$
 and subsequently the gravitational mass can be
determined by $M_{\rm s,g}=M_{\rm s,b}\left(1+{3GM_{\rm s,b}/
5R_{\rm s}c^2}\right)^{-1}$ (Dai \& Liu 2012). The corresponding
changes in the radius and inertial moment of the NS depend on the
specific equation of state. For simplicity, we use a constant NS
radius in our calculation, in view that the total mass of the accretion
disk is relatively small. On the other hand, in principle, the
material fallen onto the NS surface could repel some open field
lines into the closed region and even bury some lines. Then, the
surface dipolar magnetic field can be reduced, which may be assumed
to follow an empirical behavior as (Taam \& van den Heuvel 1986;
Shibazaki et al. 1989; Fu \& Li 2013)
\begin{eqnarray}
B_{\rm s}(t)={{B_{\rm s,i}}\over {1+ M_{\rm acc}(t)/ M_{\rm
c}}},\label{Bt}
\end{eqnarray}
where $B_{\rm s,i}$ is the initial strength of the surface dipolar
field, $ M_{\rm acc}(t)=\int_{0}^{t} \dot{M}_{\rm acc}dt'$ is the
total mass of the accreted material, and the critical mass $M_{\rm
c}$ is considered to be on the order of $(10^{-5}-10^{-3})M_{\odot}$
although it is very uncertain (Shibazaki 1989; Cheng \& Zhang 1998;
Zhang \& Kojima 2006). Tentatively, a relatively high value of
$M_{\rm c}=0.001M_{\odot}$ is adopted in our calculation. After the
surface dipolar field is suppressed, the NS would exhibit as a
normal pulsar rather than a magnetar, from the perspective of its MD
radiation. Here it should be mentioned that the surface multi-polar
magnetic fields as well as the magnetic fields internal to the
neutron star should not be significantly affected by accretion and
should remain strong as before. In any case, through Ohmic diffusion
and Hall drift, the repelled/buried field lines could finally return
to their original positions (Geppert et al. 1999) and then the NS
can become a magnetar again. However, since these diffusions are
usually very slow, the corresponding timescales could be as long as
a few hundred years (Fu \& Li 2013), which is much longer than the
time of interest here.

Accompanying with the evolution of the NS, the mass and the angular
momentum of the accretion disk evolves as
\begin{eqnarray}
{dM_{\rm disk}\over dt}&=&-\dot{M}_{\rm disk} + \dot{M}_{\rm
rec},\label{dMd}\\
 {dJ_{\rm disk}\over dt}&=& -T_{\rm disk}+T_{\rm rec}\label{dJd}.
\end{eqnarray}
Here the mass flow rate of the disk can be determined by $\dot{M}_{\rm
disk}={M_{\rm disk} / \tau_{\rm v}}$, where $M_{\rm disk}$ is the
total mass of the disk and $\tau_{\rm v}$ is the viscosity timescale.
This viscosity timescale can be given as $\tau_{\rm v}= {R_{\rm
out}^{2}/ \nu}\approx{2/ \alpha \Omega_{\rm K,out}}$, where $R_{\rm
out}$ is the outer radius of the disk and $\alpha$ is the dimensionless
viscosity parameter
 (Shakura \& Syunyaev 1973). Following Kumar et al. (2008a,b), Metzger et al. (2008), and Lei et
al. (2017), we assume that the disk has a simple ring shape and its
mass is concentrated at the outer radius. Then, the angular momentum
of the disk can be expressed by $J_{\rm disk}\simeq\sqrt{GM_{\rm
s}R_{\rm out}}M_{\rm disk}$. Finally, we can simply take
$\dot{M}_{\rm rec}=\dot{M}_{\rm pro}$ and $T_{\rm rec}=T_{\rm pro}$
for the propeller-recycling state.

\begin{figure}
\centering\resizebox{0.99\hsize}{!}{\includegraphics{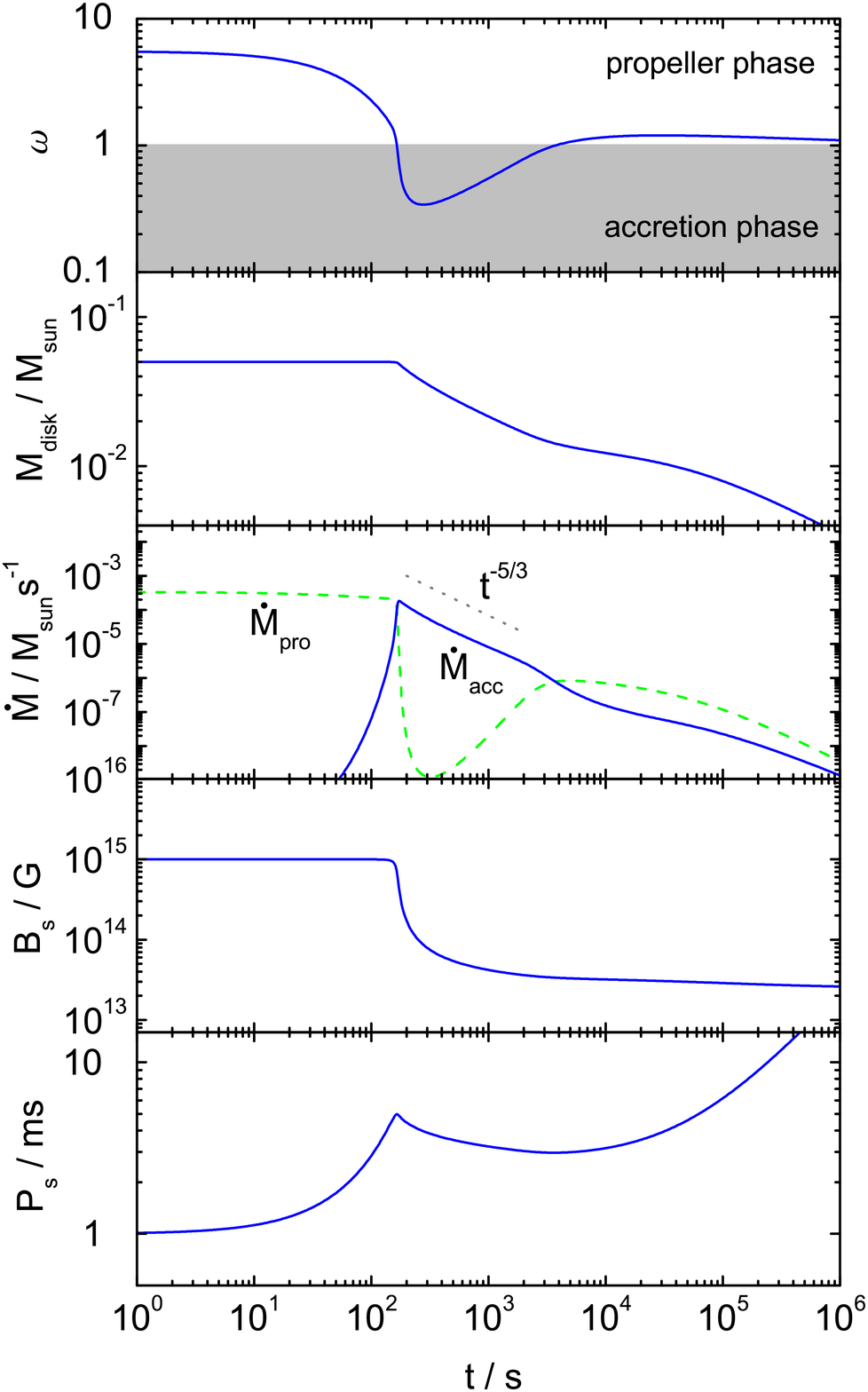}}
\caption{The co-evolution of the fundamental parameters of the NS
and the accretion disk with initial parameter values of $P_{\rm
s,i}=1{\rm ms}$, $B_{\rm s,i}=10^{15}{\rm G}$, $R_{\rm s}=12 \rm
km$, $M_{\rm s,i}=1.4 M_{\odot}$ and $M_{\rm disk,i}=0.05
M_{\odot}$. Meanwhile, the initial angular momentum of the disk is
taken as $J_{\rm disk,i}
 =5 J_{\rm s,i}$, and typical values of $\varepsilon=0.001$, $\alpha=0.05$ and $M_{\rm c}=0.001 M_{\odot}$ are adopted.} \label{Fig-evolution}
\end{figure}

By solving Equations (\ref{NSangul}), (\ref{dMd}) and (\ref{dJd}),
we present the temporal evolutions of the fundamental parameters of
the NS and the disk in Figure \ref{Fig-evolution}, where the initial
values of the parameters are taken to ensure that the propeller
condition given in Equation (\ref{Mdotsuit2}) can be satisfied at
the beginning. Specifically, as the NS spins down initially due to
the propeller, the stellar angular momentum can be effectively
transferred to the disk, although it could still be insignificant
compared with the original angular momentum of the disk. Meanwhile,
the mass of the disk also keeps invariant, because the propeller
material is assumed to be totally recycled back to the disk.
Nevertheless, as the NS continuously spins down, the disk would
finally enter the accretion phase. The accretion onto the NS surface
suppresses the surface magnetic field and thus the Alfv\'{e}n radius
is decreased, which reduces the value of $\omega$. Then, the
decrease of $\omega$ can further make the accretion more efficient.
As a result, the disk evolves from the propeller phase to the
accretion phase very quickly. Simultaneously, due to the sharp decay
of the magnetic field, the MD radiation of the NS can be suppressed
significantly. This feature can be used to explain the sharp decay
following the so-called internal-plateau GRB afterglow, which was
usually explained by the collapse of the NS (e.g., Troja et al.
2007; Lyons et al. 2010; Zhang 2014; L{\"u} \& Zhang 2014; L{\"u} et
al. 2015). After this sharp decay, the spin-down of the NS can be
sometimes dominated by the GW radiation, if the NS is deformed
significantly by its internal magnetic field (e.g., Gao et al. 2016;
Yu et al. 2018). As the accretion continues, the mass of the disk
decreases substantially, which leads to the increase of the outer
radius and as well as the viscosity timescale. Consequently, the
accretion rate can be reduced following an usual temporal behavior
of $t^{-5/3}$ (Michel 1988; Chevalier 1989). As the decrease of the
accretion rate, the value of $\omega$ can finally rebound to
increase and then, for $\omega\approx 1$, a general balance appears
between the accretion and the propeller. At this stage, the decrease
of the accretion rate would become somewhat slower than $t^{-5/3}$.

\begin{figure}
\centering\resizebox{0.99\hsize}{!}{\includegraphics{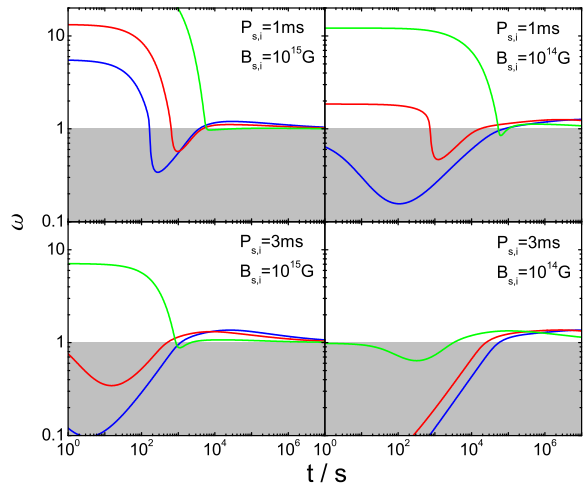}}
\caption{ The evolution of $\omega$ for different initial values of
the NS and disk parameters of $B_{\rm s,i}$, $P_{\rm s,i}$, and
$M_{\rm disk,i}$. The blue, red, and green lines correspond to disk
masses of $0.05M_{\odot}$, $0.03M_{\odot}$, and $0.01M_{\odot}$,
respectively. The other parameters are the same as in Figure 1.
 }\label{Fig-parameters}
\end{figure}

In Figure \ref{Fig-parameters}, we present the influence of the
variation of the model parameters $B_{\rm s,i}$, $P_{\rm s,i}$, and
${M}_{\rm disk,i}$ on the evolution of the crucial parameter
$\omega$, which delineates the evolutional behavior of the NS and
the disk. Obviously, as usually known, the propeller is more likely
to happen for more rapid rotation and higher magnetic fields. For a
sufficiently small disk mass to satisfy $\omega_{\rm i}>1$, the
duration of the propeller phase and the lifetime of the quasi-steady
propeller-recycling disk can be calculated by
\begin{eqnarray}
t_{\rm disk}&\approx& {(\omega_{\rm i}-1) I_{\rm s,i}\Omega_{\rm
K,m,i}\over T_{\rm pro,i}}
 = {(\omega_{\rm i}-1)GM_{\rm s,i}I_{\rm s,i}\over\Omega_{\rm
s,i}\mu_{\rm i}^{2}}\nonumber\\
&=&3.3\times10^{3}(\omega_{\rm i}-1)M_{\rm s,i,1.4}^2 R_{\rm
s,6}^{-4}B_{\rm s,i,14}^{-2}P_{\rm s,i,-3}\,\rm s. \label{tprop}
\end{eqnarray}
Nevertheless, if the magnetic field is extremely high, the MD
radiation of the NS could initially exceed the energy release due to
the propeller effect. In this case, the lifetime of the
propeller-recycling disk should be determined by the MD spin-down
timescale that is shorter than the estimate given in Equation
(\ref{tprop}). The value of $t_{\rm disk}$ can further be shortened
if the NS can collapse into a black hole before the propeller ends.
The collapse could be due to the spin-down of the NS, if the mass of
the NS is higher than the maximum mass of non-rotating NSs.
Additionally, the collapse can also happen in the accretion phase,
if accretion makes the NS's mass exceed its maximum value. No matter
whether the collapse happens or not, the eventual accretion onto the
compact object can always take place, as long as the fallback
accretion disk exists.

\section{Summary and discussion}

For a newborn NS, its early evolution could be influenced by
fallback accretion. It is found that, for a millisecond magnetar,
the accretion disk could be initially in a propeller-recycling state
for about hundreds or thousands of seconds. Nevertheless, as a result of the NS's spin-down due to the propeller
effect, the disk can eventually enter the accretion phase. During this stage, the
decrease of the accretion rate can be firstly well described by the usual temporal behavior of $t^{-5/3}$ and subsequently become somewhat slower. In any case, due
to the accretion, the surface dipolar magnetic field of the NS could
be effectively reduced and then the NS will appear as a low-field
magnetar at later time.

Such an evolutional behavior of the newborn NS is consistent with
the evidence of a possible long-lived neutron star engine in some
short GRBs (Yu et al. 2018; Li et al. 2018). On one hand, the
post-GRB NSs are strongly inferred to be mangetars with a surface
dipolar magnetic field of $\sim10^{14}-10^{15}$ G, in order to
account for the early afterglow emission of GRBs including the
plateau or flare features (Yu et al. 2010; Zhang 2013; Rowlinson et
al. 2013). On the other hand, however, the luminosity of the optical
transient emission (i.e., AT2017gfo) in the GW170817/GRB 170817A
event stringently constrained the surface dipolar magnetic field of
the remnant NS to be lower than $\sim10^{12}$ G (Ai et al. 2018; Yu
et al. 2018; Li et al. 2018). It is indicated that the MD radiation
of this remnant NS had probably been suppressed at a time of several
hours after the birth of the NS. The calculation results presented
in this paper suggest that this suppression of the MD radiation
could be caused by the delayed
 accretion from a propeller-recycling disk. The total mass of the accretion disk
is required to be not higher than several hundredths of solar mass, which is
roughly consistent with the order of magnitude of the ejecta mass of
double NS mergers and white dwarf AICs.

Finally, in principle, the disk accretion could sometimes lead to an
accretion feedback outflow, which can provide extra energy injection
to the transient emission. However, in observations, such a sharp
energy pulse at hundreds or thousands of seconds seems not
ubiquitous in the GRB afterglow emission. So, it may suggest that
the efficiency of this direct energy release due to the delayed
accretion could not be very high, e.g., lower than $\sim 10^{-4}$.
The effect of the fallback accretion is primarily reflected by its
important influence on the early evolution of the newborn NS.

\acknowledgements This work is supported by the National Natural
Science Foundation of China (Grant Nos. 11822302 and 11833003) and
the Fundamental Research Funds for the Central Universities (Grant
No. CCNU18ZDPY06).

\end{document}